\documentclass[apj,numberedappendix]{emulateapj}
\usepackage{apjfonts}

\newcommand{\be}{\begin{eqnarray}}
\newcommand{\ee}{\end{eqnarray}}

\newcommand{\lp}{\left(}
\newcommand{\rp}{\right)}

\begin{document}

\normalsize

\slugcomment{Submitted for publication in The Astrophysical Journal}


\shorttitle{Reconciling $^{56}$Ni Production in SNe Ia}
\shortauthors{Piro, A. L., Thompson, T. A., \& Kochanek, C. S.}

\title{Reconciling $^{56}$Ni Production in Type Ia Supernovae with Double Degenerate Scenarios}

\author{Anthony L. Piro\altaffilmark{1}, Todd A. Thompson\altaffilmark{2}, and Christopher S. Kochanek\altaffilmark{2}}

\altaffiltext{1}{Theoretical Astrophysics, California Institute of Technology, 1200 E California Blvd., M/C 350-17, Pasadena, CA 91125; piro@caltech.edu}

\altaffiltext{2}{Department of Astronomy and Center for Cosmology \& Astro-Particle Physics, The Ohio State University, Columbus, OH 43210, USA}


\begin{abstract}

Binary white dwarf (WD) coalescences driven by gravitational waves or collisions in triple systems are potential progenitors of Type Ia supernovae (SNe Ia). We combine the distribution of $^{56}$Ni inferred from observations of SNe Ia with the results of both sub-Chandrasekhar detonation models and direct collision calculations to estimate what mass WDs should be exploding in each scenario to reproduce the observations.  These WD mass distributions are then compared with the observed Galactic WD mass distribution and Monte Carlo simulations of WD-WD binary populations.  For collisions, we find that the average mass of the individual components of the WD-WD binary must be peaked at $\approx0.75\,M_\odot$, significantly higher than the average WD mass in binaries or in the field of $\approx0.55-0.60\,M_\odot$. Thus, if collisions produce a large fraction of SNe Ia, then a mechanism must exist that favors large mass WDs. On the other hand, in an old stellar population, collisions would naturally result in a class of low luminosity SNe Ia, and we suggest these may be related to 1991bg-like events. For sub-Chandrasekhar detonations, we find that the average mass of the exploding WDs must be peaked at $\approx1.1\,M_\odot$. This is interestingly similar to the average sum of the masses in WD-WD binaries, but it is not clear (and should be further explored) whether double degenerate mergers would be sufficiently efficient at synthesizing $^{56}$Ni to match the observed yields. If not, then actual $\approx1.1\,M_\odot$ WDs would be needed for sub-Chandrasekhar detonations. Since such high mass WDs are produced relatively quickly in comparison to the age of the environments where SNe Ia are found, this would require either accretion onto lower mass WDs prior to ignition or a long timescale between formation of the $\approx1.1\,M_\odot$ WD and ignition (such as set by gravitational wave emission or binary interactions).
\end{abstract}



\section{Introduction}
\label{sec:introduction}

The use of Type Ia supernovae (SNe Ia) as precision probes of cosmology \citep[e.g.,][]{rie98,per99} will ultimately be limited by systematic uncertainties. Understanding and minimizing these uncertainties should be advanced by having a complete physical understanding of the underlying mechanism behind the explosion. Thus, one of the consequences of the focus on SNe Ia as cosmological distance indicators has been to emphasize the enormous theoretical uncertainties that remain about these events.

It is generally accepted that SNe Ia result from unstable thermonuclear ignition of degenerate matter \citep{hf60} in a C/O white dwarf (WD), but, frustratingly, the specific progenitor systems have not yet been identified. The three main candidates are (1) stable accretion from a non-degenerate binary companion until the Chandrasekhar limit is reached \citep[single degenerates,][]{wi73}, (2) the merger of two C/O WDs \citep[double degenerates,][]{it84,web84}, or (3) accretion and detonation of a helium shell on a C/O WD that leads to a prompt detonation of the core \citep[double detonations,][]{ww94,la95}. An important outstanding problem is to understand how these scenarios contribute to the SNe Ia we observe, and whether any one channel is dominant.

In recent years, the double degenerate mechanism has been increasingly at the center of attention. Observationally, there are arguments in favor of this scenario from the non-detection of a companion in pre-explosion imaging of nearby SNe Ia \citep{Li et al. 2011a}, the lack of radio emission from SNe Ia \citep{Hancock et al. 2011,Horesh et al. 2012}, the lack of hydrogen emission in nebular spectra of SNe Ia \citep{Leonard 2007,Shappee et al. 2013a}, a lack of a signature of ejecta interaction with a companion \citep{Kasen 2010,Hayden et al. 2010,Bloom et al. 2012}, and the missing companions in SNe Ia remnants \citep{Schaefer Pagnotta 2012} even though they should be super-luminous \citep{Shappee et al. 2013b}. In addition, the delay time distribution of SNe Ia follow a power-law distribution as is expected for double degenerates \citep{Maoz et al. 2010,Graur et al. 2011,Barbary et al. 2012,Sand et al. 2012}. Potential problems with matching the rate of SNe Ia with double degenerate mergers may be alleviated if the mergers are in sub-Chandrasekhar WD-WD binaries \citep{van Kerkwijk et al. 2010,Badenes Maoz 2012}.

On the theoretical side, double degenerate scenarios have historically been disfavored because accretion after tidal disruption triggers burning that turns the C/O WD into a O/Ne WD \citep{Nomoto Iben 1985, Saio Nomoto 1998}, which then collapses to a neutron star due to electron captures \citep{Nomoto Kondo 1991}. This problem remains even with more detailed treatments of the long-term evolution of the merger remnant \citep{Shen et al. 2012}. More recently though, the double degenerate scenario has been revitalized by new simulations which indicate that ignition may be triggered by a detonation in an accretion stream \citep{gul10,Dan et al. 2012} or in ``violent mergers'' involving massive WDs \citep{Pakmor et al. 2012}. WDs may also explode in direct collisions \citep{Rosswog et al. 2009,Raskin et al. 2010,Kushnir et al. 2013}, which would be another way for double degenerates to give rise to SNe Ia. While this scenario may have been viewed as unlikely only a few years ago, it is now reasonably clear that triple systems are more common \citep{Raghavan et al. 2010} and that the Kozai mechanism both greatly accelerates binary mergers \citep{Thompson 2011} and drives direct collisions \citep{Katz Dong 2012} in such systems.

With this increased focus on double degenerate scenarios, the time is ripe to make better comparisons to observed and theoretical populations of WD binaries. In the present work we investigate this problem using the following strategy. First, the observed  luminosity distribution of SNe Ia implies a corresponding distribution of radioactive $^{56}$Ni synthesized, which we present in \S \ref{sec:m56}. Next, the relation between WD mass and $^{56}$Ni yield for a given explosion scenario means that certain mass WDs much be exploding to produce the SNe Ia that we observe. In \S \ref{sec:compare}, we perform this exercise and find that sub-Chandrasekhar detonation models and collision calculations favor the explosion of $\approx1.1\,M_\odot$ and $\approx0.75\,M_\odot$ WDs, respectively. The implications of this conclusion are then investigated with comparisons to the mass distribution of field WDs and Monte Carlo calculations of WD-WD binaries in \S \ref{sec:binary}. We conclude in \S \ref{sec:conclusion} with a summary of our results and a discussion of future explorations of this problem.

\section{The Observed $^{56}$Ni Distribution}
\label{sec:m56}

We begin by investigating the range of $^{56}$Ni masses, $M_{56}$, produced in SNe Ia. To do this we focus on the volume-limited sample of 74 SNe Ia within $80\,{\rm Mpc}$ from the Lick Observatory Supernova Search \citep[LOSS,][]{Li et al. 2011b}. The sample is estimated to be 98\% complete due to the high peak luminosity of these SNe. There may be some bias because LOSS targets specific galaxies rather than broadly surveying the sky. For example, the sample is mostly composed of normal SNe Ia, without any super-Chandrasekhar events \citep[e.g., SN 2003fg,][]{Howell et al. 2006} possibly because these tend to be associated with low-metallicity dwarf galaxies \citep{Khan et al. 2011} that are not a focus of the survey.

By combining modeling of the late-time nebular spectra of SNe Ia with measurements of their bolometric peak, \citet{Stritzinger et al. 2006} demonstrated that $\Delta m_{15}(B)$ (the $B$-band magnitude change $15\,{\rm days}$ post peak) is a reliable indicator of the $^{56}$Ni yield. This has the additional advantage that it is relatively insensitive to extinction corrections in comparison to other possible $^{56}$Ni indicators\footnote{We thank S. Dong for bringing this to our attention, so that we could correct our $^{56}$Ni yields from a previous version of this manuscript.}. Therefore to infer the $^{56}$Ni mass produced in each SN Ia, we use the decline rate-nickel mass relation presented in \citet{Mazzali et al. 2007},
\be
	M_{56}/M_\odot = 1.34 - 0.67\Delta m_{15}(B),
	\label{eq:m15}
\ee
which has an rms dispersion of $0.13\,M_\odot$. Unfortunately, $\Delta m_{15}(B)$ is not available directly in \citet{Li et al. 2011b}, and so we compiled a list of $\Delta m_{15}(B)$ values from a number of other references \citep{Krisciunas et al. 2000,Krisciunas et al. 2004,Modjaz et al. 2001,Hicken et al. 2009,Wang et al. 2009,Ganeshalingam et al. 2010, Ganeshalingam et al. 2012,Folatelli et al. 2013,Foley et al. 2013}. This exercise still left 14 out of the 74 SNe Ia without measured $\Delta m_{15}(B)$ values. Out of these, 8 were completely normal SNe Ia (7 of which were in late-type galaxies), and thus we assume they produce $M_{56}\approx0.55-0.65\,M_\odot$, consistent with all their other properties. This assumption did not change the overall $^{56}$Ni distribution we derived appreciably. The other 6 were all 1991bg-like SNe Ia, all of which were in early-type galaxies. It has been well established that this subluminous class of SNe Ia synthesize a small amount of $^{56}$Ni \citep{Sullivan et al. 2011}, and thus we assume that each of these SNe have $M_{56}=0.1\,M_\odot$, consistent with other members of this class. Our assumptions for these objects indeed made a noticeable difference in the derived $^{56}$Ni distribution, which we discuss later in this section.

\begin{figure}
\epsscale{1.2}
\plotone{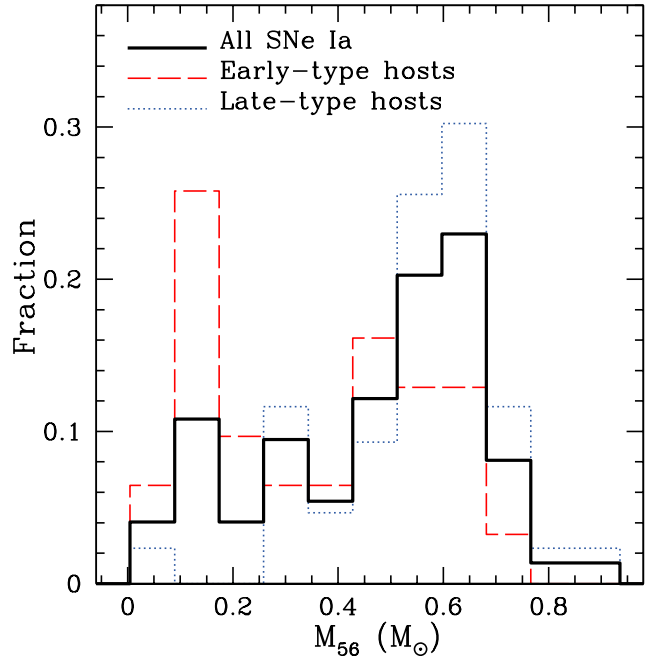}
\caption{Histograms showing the fraction of SNe Ia that produce different amounts of $^{56}$Ni found by combining the volume-limited LOSS sample of SNe Ia \citep{Li et al. 2011b} with the decline rate-nickel mass relation \citep{Mazzali et al. 2007}.}
\label{fig:m56}
\epsscale{1.0}
\end{figure}

In Figure \ref{fig:m56} we plot histograms summarizing this analysis. We compare all SNe Ia (black, solid line) with SNe Ia from early-type host galaxies (red, dashed line) and late-type host galaxies (blue, dotted line). The overall peak is at $M_{56}\approx0.60\,M_\odot$, as has been well-established for typical SNe~Ia. It is also well-known that SNe~Ia are on average brighter in late-type galaxies in comparison to early-type galaxies \citep[e.g.,][]{Howell et al. 2007}, which corresponds to the average SNe Ia in a late-type galaxy producing $\approx0.13\,M_\odot$ more $^{56}$Ni \citep{Piro Bildsten 2008}. This difference is also apparent in Figure \ref{fig:m56}. Because combining both types of hosts provides the best statistics on the $M_{56}$ distribution, we focus on the overall luminosity distribution of all SNe Ia together for most of the remainder of the present study. In the future, similar analysis can and should be applied to SNe~Ia with early- and late-type hosts separately.

An additional feature of Figure \ref{fig:m56} that deserves mention is the apparent peak in the $^{56}$Ni production at around \mbox{$M_{56}\approx0.1\,M_\odot$,} which is especially conspicuous for early-type galaxies. This is due exclusively to the 1991bg-like events\footnote{We note that it is possible that many of the events that we assumed produced $M_{56}\approx0.1\,M_\odot$ could have just as well produced $M_{56}\approx0.2\,M_\odot$ \citep[as inferred in][]{Gonzalez et al. 2012}, and this peak would have been just as prominent.}. Although it has long been appreciated that this subluminous class of SNe Ia is distinct in many ways, their contribution to the overall SNe Ia rate in a volume limited sample is dramatic. Out of 31 SNe Ia in early-type galaxies in the LOSS sample, 10 are 1991bg-like, which is more than $30\%$\footnote{In fact, even among {\em all} nearby galaxies, the subluminous SN Ia rate has been estimated to be $\sim15-30\%$ of all SNe Ia \citep{Li et al. 2011b,Gonzalez et al. 2011}.}. In comparison, only a single 1991bg-like event occurred in a late-type galaxy. Furthermore, the subluminous SN Ia rate is found to be consistent with only being dependent on the galactic mass \citep[as opposed to depending on the star formation rate,][]{Gonzalez et al. 2011}. Clearly an old stellar population is a crucial prerequisite for producing this class of SNe, which is a point we explore further in \S \ref{sec:collisions}.

\section{Progenitor White Dwarf Mass Distributions}
\label{sec:compare}

Different SNe Ia mechanisms imply different relations between the mass of the exploding WD and the amount of $^{56}$Ni synthesized. For the present work we focus on two scenarios for double degenerate explosions as follows.

\begin{figure}
\epsscale{1.2}
\plotone{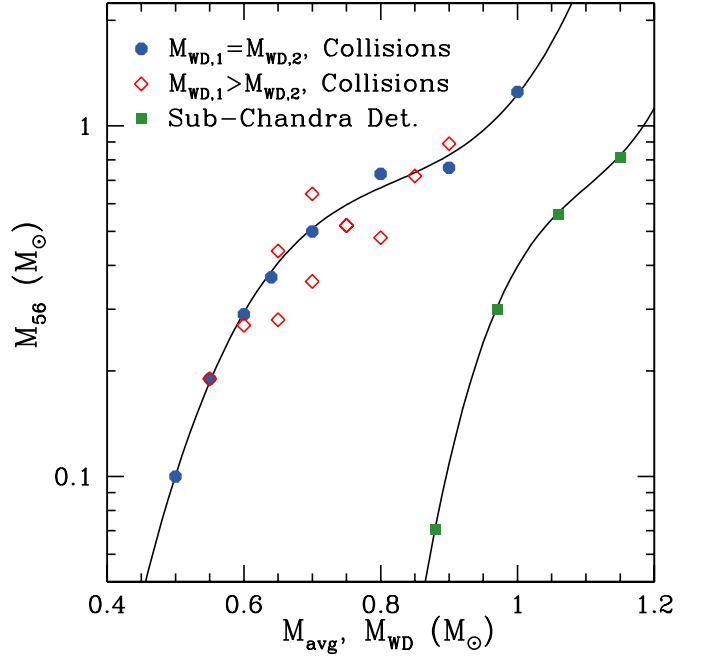}
\caption{Mass of $^{56}$Ni produced for equal mass collisions (blue, filled circles), non-equal mass collisions (red, open diamonds), and sub-Chandrasekhar detonations (green, filled squares). The collision results are taken from the high resolution simulations of  \citep{Kushnir et al. 2013} and are plotted against the average mass of the two colliding WDs. The sub-Chandrasekhar detonations are taken from \citep{Sim et al. 2010} and are plotted against the mass of the single exploding WD. The solid lines are the fits summarized in equations (\ref{eq:sim_yield}) and (\ref{eq:kushnir_yield}).}
\label{fig:ni_yield}
\epsscale{1.0}
\end{figure}

{\bf Sub-Chandrasekhar Detonations:} We use the work of \citet{Sim et al. 2010}, which considers the detonation of sub-Chandrasekhar WDs. They find that they can reproduce the range of $M_{56}$ needed for the observed typical SNe Ia given a relatively narrow spread of WD masses of $M_{\rm WD}\approx 0.97-1.15\,M_\odot$. Although they do not study a specific mechanism for triggering these detonations, such an event could occur in a double detonation following helium accretion from a non-degenerate helium star or a helium WD \citep{Fink et al. 2010} or in a WD-WD merger from a circular orbit \citep{van Kerkwijk et al. 2010}. We fit their results with a third-order polynomial,
\be
	\log_{10}(M_{56}/M_\odot)
	= &&56.47(M_{\rm WD}/M_\odot)^3-186.30(M_{\rm WD}/M_\odot)^2
	\nonumber
	\\
	&&+206.56(M_{\rm WD}/M_\odot)-77.13,
	\label{eq:sim_yield}
\ee
to estimate the $^{56}$Ni as a function of the detonating WD mass. The large number of digits in each of these coefficients is not meant to represent the significant figures of the $^{56}$Ni yield estimation, but merely a consequence of making an accurate fit when using a third-order polynomial. This fit is plotted in Figure \ref{fig:ni_yield} in comparison to the \citet{Sim et al. 2010} $^{56}$Ni yields (green, filled squares).

{\bf Collisions:} Another promising way to ignite detonations in double degenerate systems is via collisions, for which we consider the calculations of \citet{Kushnir et al. 2013}. They generally find that the $^{56}$Ni yield only depends on the average mass of the constituents in the collision,
\be
	M_{\rm avg} = 0.5(M_{{\rm WD},1}+M_{{\rm WD},2}),
	\label{eq:mavg}
\ee
where $M_{{\rm WD},1}$ and $M_{{\rm WD},2}$ are the primary and secondary masses of the WDs that are colliding, respectively. Again, we fit their $^{56}$Ni yield with a third-order polynomial,
\be
	\log_{10}(M_{56}/M_\odot)
	= &&16.92(M_{\rm avg}/M_\odot)^3-41.73(M_{\rm avg}/M_\odot)^2
	\nonumber
	\\
	&&+35.16(M_{\rm avg}/M_\odot)-10.26.
	\label{eq:kushnir_yield}
\ee
This fit is plotted in Figure \ref{fig:ni_yield} in comparison to the \citet{Kushnir et al. 2013} $^{56}$Ni yields for equal mass collisions (blue, filled circles) and non-equal mass collisions (red, open diamonds), where we only use their results from high resolution simulations (see their Table 1). In the future, a more complete comparison with collision calculations should also include the mass ratio and impact parameter of the collision. For example, in the best-resolved smooth particle hydrodynamic 3D simulations of \citet{Raskin et al. 2010}, they generally find $\sim10\%$ more $^{56}$Ni production in equal mass head-on collisions in comparison to \citet{Kushnir et al. 2013}, and a significant decrease in $^{56}$Ni for unequal mass head-on collisions. For the time being, we delay doing a comparison with this other set of calculations until there exists a more complete survey over the full range of parameters.

We combine the $^{56}$Ni distribution in Figure \ref{fig:m56} with $M_{56}$ yields from equations (\ref{eq:sim_yield}) and (\ref{eq:kushnir_yield}) to derive the WD mass distribution needed to reproduce the observations in the sub-Chandrasekhar detonation and collision scenarios, respectively.  The results are shown in Figure \ref{fig:kushnir_wds} (red dashed and blue dotted lines, respectively, both shaded) together with the mass distribution of Galactic field WDs (black, solid line), which we discuss in the following section. Figure \ref{fig:kushnir_wds} demonstrates that collisions must come from WD-WD binaries with component masses of \mbox{$M_{\rm avg}\approx0.75\,M_\odot$} in order to reproduce the observed SNe Ia luminosity function, whereas sub-Chandrasekhar detonations must come from WDs that are exploding with masses of $\approx1.1\,M_\odot$. Thus if one of these channels is the dominant mechanism for producing SNe Ia, then there must be a reason why this corresponding WD mass is preferentially exploding. In the following sections, we discuss the implications of these mass distributions and investigate what constraints they allow us to place on the relation of these scenarios to the observed SNe~Ia. 

\subsection{Comparisons to Field White Dwarfs}
\label{sec:field}

\begin{figure}
\epsscale{1.2}
\plotone{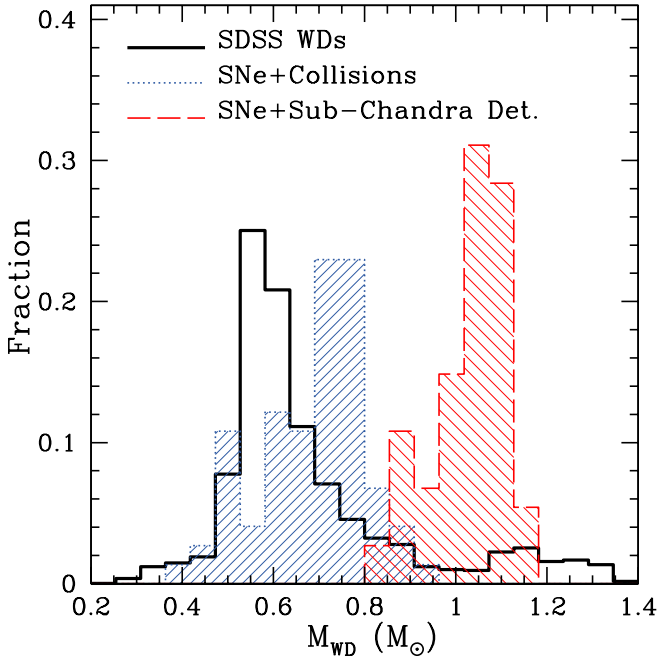}
\caption{Histograms of the distribution of WD masses $M_{\rm WD}$ from the SDSS WD catalog \citep[][black solid line]{Kepler et al. 2007} as compared to the WD masses needed for sub-Chandrasekhar detonations \citep[][red dashed line, lightly shaded histogram]{Sim et al. 2010} and average collision masses $M_{\rm avg}$ needed for head-on collisions \citep[][blue dotted line, darkly shaded histogram]{Kushnir et al. 2013}. }
\label{fig:kushnir_wds}
\epsscale{1.0}
\end{figure}

We next compare these inferred mass distributions with the volume-corrected mass distribution of spectroscopically confirmed WDs from Sloan Digital Sky Survey Data Release 4 \citep{Kepler et al. 2007}. Summing all DA and DB WDs, the total sample contains over 1,800 WDs. Plotting the corresponding histogram in Figure \ref{fig:kushnir_wds} (black, solid line), the average mass of field WDs is $\approx0.55-0.60\,M_\odot$, and it is clearly different than either the sub-Chandrasekhar detonation or collision scenarios. In particular, this comparison shows that collisions between average-mass WDs of $\sim0.6\,M_\odot$ produce too little $^{56}$Ni to power the average observed SNe Ia.  Thus, if collisions are responsible for the majority of SNe Ia that we see, they must pick out high-mass progenitors and collisions must be suppressed in binaries with average-mass WD constituents.

Although the mass distribution inferred for sub-Chandrasekhar detonations is also inconsistent with the overall field WD population, as one would expect naively, its peak at $\approx1.1\,M_\odot$ is not too dissimilar from the secondary high-mass peak in the field WD population at $\approx1.2\,M_\odot$. It has been suggested that the high mass peak is due to mergers of lower mass WDs \citep{Vennes 1999,Liebert et al. 2005}, which may indicate a connection between mergers and sub-Chandrasekhar detonations. The implication may be that either (1) SN Ia progenitors are coming from the same binary mergers that would produce these massive WDs or that (2) the WDs merged first and then the explosion was triggered later, as in a double detonation. In the first case, it is unclear why some WDs would explode upon merger (producing SNe Ia) while other WDs would produce the massive field WDs. In the second case, it seems like a specialized set of circumstances would be needed to first produce massive WD via a merger and then have an event that subsequently triggered an explosion\footnote{Later we discuss scenarios that have been explored in population synthesis calculations which may indeed allow the WD to accrete and become more massive before unstably igniting, as is needed for this scenario.}. On the other hand, it has also been argued that the kinematics of massive WDs are consistent with single star evolution \citep{Wegg Phinney 2012} rather than being the product of mergers. The suggestion is then that perhaps SNe Ia come from more massive WDs that are simply the result of more massive main sequence stars. Whatever the conclusion is, the rough similarity of these peaks clearly requires more investigation, some of which we conduct in the next section.

\section{$^{56}$Ni Yields from Binary Populations}
\label{sec:binary}

So far we have made comparisons to field WDs, but SN~Ia progenitors are expected to be in binary (or perhaps triple) systems. We assess the impact of binarity with a Monte Carlo binary mass distribution calculation. Instead of performing a detailed population synthesis \citep[e.g.,][and references therein]{Belcynski et al. 2008} we use a simpler model to focus on certain generic aspects of WD-WD binary populations in the absence of mass transfer and binary interactions. This allows us to estimate the average and total mass in WD-WD binaries for comparisons with explosion scenarios, and to explore the impact of age and star formation history on the expected $^{56}$Ni yields.

Our analysis proceeds as follows. First, we consider a distribution of main sequence stars with mass $M_1$, which obeys a Salpeter initial mass function,
\be
	dN/dM_1 \propto M_1^{-2.35}.
\ee
Next we consider companion masses $M_2$, which are assigned a flat distribution in mass so that the probability $P(q)$ is constant, where $q=M_2/M_1\le1$. For a given binary we can evaluate the final masses of each of the WDs that are created using the initial mass-final mass relation \citep{Kalirai et al. 2008},
\be
	M_{{\rm WD},i}/M_\odot = 0.109 M_i +  0.394.
\ee
We assume a maximum mass of $7\,M_\odot$ for $M_1$ and $M_2$ to produce a C/O WD. The lower mass limit is taken to be $0.9\,M_\odot$ so as to focus on progenitors of C/O WDs rather than helium WDs. The timescale for formation of a double degenerate binary is
\be
	t_{\rm form} = t_{\rm birth} + 10\lp\frac{M_2}{M_\odot}  \rp^{-2.5}{\rm Gyr},
\ee
where $t_{\rm birth}$ is the time when the main-sequence binary was first created. Note that $t_{\rm form}$ is controlled by mass $M_2$, since the lower mass secondary takes longer to evolve off the main sequence. Finally, there is an explosion time given by the sum of the formation time and the timescale for ignition of a detonation or a collision,
\be
	t_{\rm exp} = t_{\rm form} + t_{\rm ign}.
\ee
Given this set of prescriptions, we can assemble a large number of WD binaries with a distribution of masses and associated timescales using Monte Carlo methods. We can then estimate the current distribution now at time $t_{\rm now}\approx 13.7\,{\rm Gyr}$ by asking which binaries have $t_{\rm now}>t_{\rm form}$ and $t_{\rm now}<t_{\rm exp}$, in other words, those binaries that have had enough time to produce double degenerates, but have not yet exploded as SNe Ia. In this way we estimate a WD-WD binary mass distribution for comparison with the sub-Chandrasekhar detonation and collision scenarios.

\subsection{$^{56}$Ni from Collisions}
\label{sec:collisions}

\begin{figure}
\epsscale{1.2}
\plotone{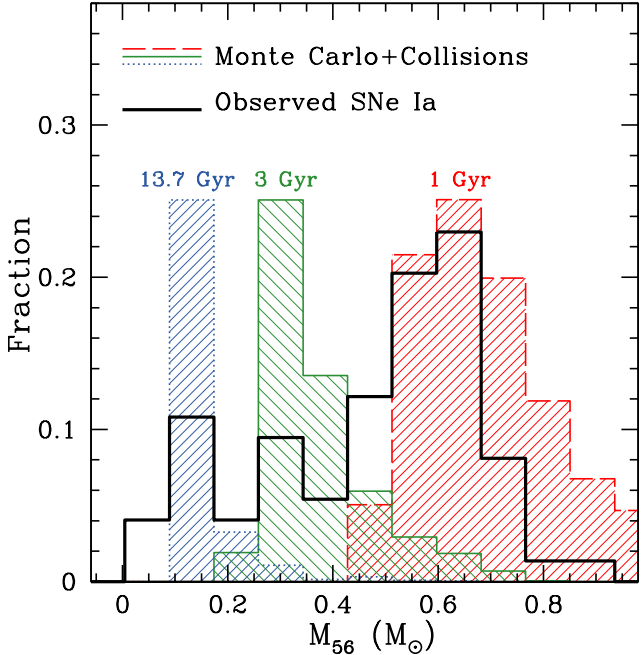}
\caption{Histograms of the distribution of $^{56}$Ni from Monte Carlo binary estimates using the collision scenario \citep{Kushnir et al. 2013}. The black solid line is the observed $^{56}$Ni distribution (from Fig. \ref{fig:m56}) and the colored histograms correspond to calculations using a burst of star formation at times of $t_{\rm burst}=13.7\,{\rm Gyr}$ (blue dotted line), $3\,{\rm Gyr}$ (green solid line), and $1\,{\rm Gyr}$ (red dashed line). The color histograms have been arbitrarily normalized to ease comparison.}
\label{fig:kushnir_mc}
\epsscale{1.0}
\end{figure}

In Figure~\ref{fig:kushnir_mc} we compare the $^{56}$Ni yield expected from our Monte Carlo calculations for collisions to the $^{56}$Ni distribution we derived from the volume-limited sample of SNe Ia as was shown in Figure \ref{fig:m56} (black, solid line). For these calculations we set $t_{\rm ign}=100\,{\rm Myr}$, although we find that the results do not depend sensitively on this assumption as long as $t_{\rm ign}\lesssim t_{\rm form}$. We focus on cases were $t_{\rm ign}$ is relatively short since this is expected for the collision scenario \citep{Katz et al. 2011}, and for sub-Chandrasekhar detonations (which will be addressed in the next section) it will allow us to assess whether the formation timescale alone is sufficient to match the observed $^{56}$Ni distribution. To set $t_{\rm birth}$, we assume a burst of star formation at some time in the past at $t_{\rm burst}$ which then lasts for $1\,{\rm Gyr}$ with a flat probability over this time. By varying $t_{\rm burst}$ we can investigate the impact of age on the resulting distribution of WD-WD binary masses. Figure~\ref{fig:kushnir_mc} plots histograms for $t_{\rm burst}=13.7\,{\rm Gyr}$ (blue, dotted line), $3\,{\rm Gyr}$ (green, solid line), and $1\,{\rm Gyr}$ (red, dashed line). Each of these histograms has been arbitrarily normalized to ease comparison with the observed distribution. Although this is a simple model, intuition about more complicated star formation histories can be gained by simply considering the integral of many of these individual star bursts.

Figure~\ref{fig:kushnir_mc} shows that to produce typical SNe Ia, collisions must occur between stars that formed rather recently, on the order of $\sim1\,{\rm Gyr}$ ago. Although this is obviously similar to our previous conclusion that high mass WDs are needed for the collision scenario to produce most SNe Ia, this comparison makes it explicit just how limiting this statement is. In investigations of SN Ia rates, there is evidence that both ``prompt'' and ``delayed'' components are needed \citep{Mannucci et al. 2005, Mannucci et al. 2006, Scannapieco Bildsten 2005, Sullivan et al. 2006}. In this context, we find that collisions can only produce normal SNe Ia as a prompt contribution. Conversely, this makes it difficult to see how collisions can produce a significant number of normal SNe Ia in a delayed component unless there is some mechanism that makes collisions between higher mass WDs more likely.

On the other hand, Figure \ref{fig:kushnir_mc} also demonstrates that for sufficiently old stellar environments (blue dotted histogram), collisions may be important for producing low luminosity SNe Ia, and indeed they have a $^{56}$Ni yield consistent with that seen from 1991bg-like events. The fact 1991bg-like SNe happen almost exclusively in early-type galaxies makes collisions an enticing explanation. It has been speculated upon before that 1991bg-like events are from WD-WD collisions, but in the context of more massive collisions that are inefficient at producing $^{56}$Ni \citep{Pakmor et al. 2010}. The problem with this hypothesis is that $t_{\rm ign}$ must be much longer than $t_{\rm form}$ to have such massive ($\sim0.9\,M_\odot$) WDs merging in old stellar environments. \citet{Pakmor et al. 2010} note this problem and speculate that the collision timescale may just naturally be long. Unfortunately, this does not explain why evidence of many more slightly lower mass collisions are not seen, since they would be favored by the initial mass function. If the $^{56}$Ni yields of \citet{Kushnir et al. 2013} are correct, then this problem is alleviated because {\em 1991bg-like events naturally match what it is expected for collisions between the most abundant mass WDs.} An important area of future research will be to investigate the expected rate of such collisions to understand whether they can be as high as the rates seen by LOSS in early-type galaxies.
\\

\subsection{$^{56}$Ni from sub-Chandrasekhar Detonations}
\label{sec:detonations}

\begin{figure}
\epsscale{1.2}
\plotone{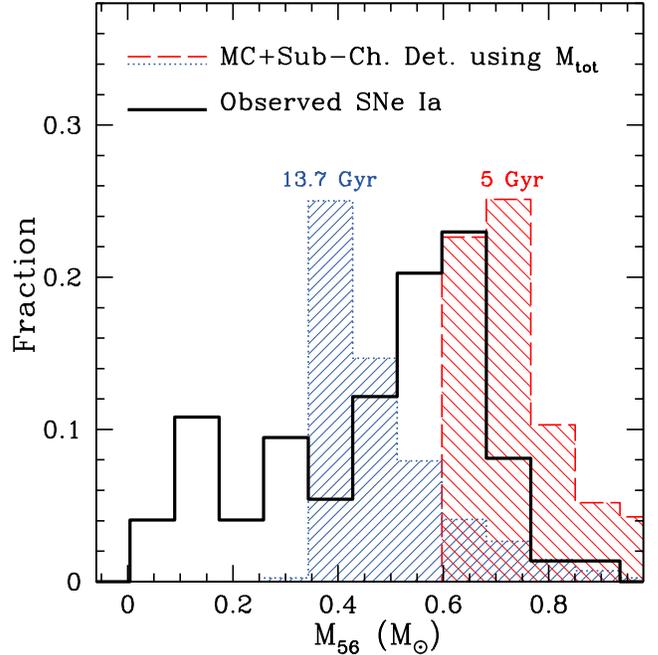}
\caption{Similar to Figure \ref{fig:kushnir_mc}, but instead for the sub-Chandrasekhar detonation scenario \citep{Sim et al. 2010}. The colored histograms correspond to calculations using a burst of star formation at times of $t_{\rm burst}=13.7\,{\rm Gyr}$ (blue dotted line) and $5\,{\rm Gyr}$ (red dashed line).}
\label{fig:sim_mc}
\epsscale{1.0}
\end{figure}

For the sub-Chandrasekhar detonation scenario, there are two potential masses we could identify for the triggering of the detonation, either (1) the primary mass or (2) the total mass of the binary,
\be
	M_{\rm tot} = M_{{\rm WD},1}+M_{{\rm WD},2}.
\ee
In the first case we know from Figure \ref{fig:kushnir_wds} that the average mass of the detonating primary must be $\approx1.1\,M_\odot$. Although it is possible that $t_{\rm ign}$ for such a primary could be sufficiently long to allow such high mass WDs to last long enough to produce typical SNe Ia in both young and old stellar environments, it is not immediately clear why a $\approx1.1\,M_\odot$ primary would be favored for explosion in comparison to, say, a $\approx1.0\,M_\odot$ primary without appealing to some sort of binary interactions. In the next section we discuss the results of population synthesis analysis which takes this into account, but for the simpler population model we are using, this physics is outside the context of what we are investigating.

So instead we focus on the latter case of using $M_{\rm tot}$ to estimate the $^{56}$Ni production. The idea here would be that a WD-WD merger could potentially be qualitatively similar to the mass budget of just combining the two WDs. One should be careful here because exploding two $0.6\,M_\odot$ WDs separately will yield much less $^{56}$Ni mass than exploding one $1.2\,M_\odot$ WD. Using $M_{\rm tot}$ corresponds to the assumption that following the merger the density reaches a configuration roughly like the larger mass object, which may require some time to adjust to the increase in mass \citep[e.g.,][]{Shen et al. 2012}. With these caveats in mind, we show the results of our Monte Carlo calculations in Figure \ref{fig:sim_mc}. In this case the average $^{56}$Ni yield seen in observations is consistent with WDs from stars that formed $t_{\rm burst}\approx5-7\,{\rm Gyr}$ ago (red, dashed line), since this is what is needed for binaries with $M_{\rm tot}\approx 1.1\,M_\odot$. We conclude from this comparison that it is at least plausible that {\em the average SNe Ia could be explained by sub-Chandrasekhar mergers as long as the total mass of the binary corresponds to the explosion mass.} On the other hand, going to especially old stellar populations (blue, dotted line) will still make a SNe Ia with a relatively normal amount of $^{56}$Ni production ($M_{56}\sim0.4\,M_\odot$), so it is difficult to explain the especially subluminous SNe Ia if the entire mass of the binary is involved in the detonation, and we limit ourselves to C/O WDs.

\section{Conclusion and Discussion}
\label{sec:conclusion}

We have conducted an initial investigation exploring the implications of the collision and sub-Chandrasekhar detonation scenarios as possible progenitors of SNe Ia. First, we derived the $^{56}$Ni distribution from a volume-limited sample of SNe Ia (Figure \ref{fig:m56}). This was used to infer the distribution of WD masses that must be exploding in each scenario in order to match the observations, and then to make comparisons with the observed field WD mass distribution  (Figure \ref{fig:kushnir_wds}). Using a simple Monte Carlo population analysis, we investigated the $^{56}$Ni yield as a function of stellar age to explore the viability of each scenario.

\subsection{Sub-Chandrasekhar Detonation Scenario}

Our main conclusion for the sub-Chandrasekhar detonation scenario \citep{Sim et al. 2010} is that it requires the explosion of WDs with an average mass of $\approx1.1\,M_\odot$. This is clearly inconsistent with the general mass distribution of {\it single} field WDs, but may be similar to a population of more massive WDs which have a distribution peak at around $\approx1.2\,M_\odot$ (Figure \ref{fig:kushnir_wds}). We then explored the $^{56}$Ni yield from populations of various ages and found that a burst of star formation at $t_{\rm burst}\approx5-7\,{\rm Gyr}$ ago would allow sub-Chrandrasekhar detonations to explain typical SNe Ia (Figure \ref{fig:sim_mc}). Although this connection is enticing, there are problems that still need to be sorted out to understand its importance. If two WDs merge, the density of the resulting WD that experiences the detonation need not be equivalent to a WD that has a mass which is the sum of the two constituents. Our analysis would therefore benefit from some conversion factor, which would give a better estimate of how much material is at a sufficiently high density to produce $^{56}$Ni. If the conversion factor is low (for example, if ignition occurs when a large fraction of the material in a merging WD-WD binary is still at relatively low densities), then it may be related to some of the lower luminosity SNe Ia that are difficult to explain with sub-Chandrasekhar detonations using our simplistic model.

Another scenario for getting a sub-Chandrasekhar detonation, which was outside the context of our simple population model (as discussed in \S \ref{sec:detonations}, this would require binary interaction physics or a long $t_{\rm ign}$), was that the primary mass could be the determining factor for estimating the $^{56}$Ni yield. \citet{Ruiter et al. 2013} considered this case, and also concluded that the average exploding WD mass much be $\approx1.1\,M_\odot$. Using population synthesis, it was found that the most promising avenue for creating such a progenitor was by taking a somewhat smaller mass WD ($\approx0.8-0.9\,M_\odot$), increasing its mass up to $\approx1.1\,M_\odot$ via helium-rich accretion from its companion, and then eventually merging with that companion \citep{Ruiter et al. 2013}. Whether or not this scenario happens robustly in detailed accretion models (for example, that there is little mass loss during helium accretion as assumed in these population synthesis calculations) requires more investigation.

Instead of a merger, yet another way to ignite the primary in a sub-Chandrasekhar detonation would be with a double detonation, where a helium-rich layer is accreted and detonated, triggering the C/O core \citep{ww94,la95}. Although in the past this mechanism has been disfavored because it produces colors and spectra that do not match normal SNe Ia \citep{Kromer et al. 2010}, more detailed treatments of the helium burning suggest that this problem may be alleviated \citep{Townsley et al. 2012,Moore et al. 2013}. Whatever the answer may be, the fact remains that $\approx1.1\,M_\odot$ WDs must somehow be favored for exploding in sub-Chandrasekhar detonations in comparison to any other mass. The results of our work emphasize the importance of this litmus test for any future similar classes of models.

\subsection{Collision Scenario}

For the collision scenario \citep{Kushnir et al. 2013} we find that the average mass of an exploding WD must be $\approx0.75\,M_\odot$. Although collisions could therefore produce typical SNe Ia in especially young environments, it is hard to see how collisions could generate a significant fraction of the normal \mbox{SNe Ia} that we observe. We note that DB WDs and magnetic WDs are generally more massive than DA and non-magnetic WDs \citep{Wickramasinghe Ferrario 2000,Kepler et al. 2007}, but  there is not a clear reason why these populations should be expected to participate in collisions more often than regular WDs. There are several ways to alleviate this inconsistency: the $^{56}$Ni yields in hydrodynamic calculations are too low by $\approx0.3\,M_\odot$ \citep[which seems unlikely given the convergence considerations in][]{Raskin et al. 2010,Kushnir et al. 2013}, the conversion from $\Delta m_{15}(B)$ to $M_{56}$ (eq. [\ref{eq:m15}]) is too high by the same factor, or the physics associated with glancing collisions that yield subsequent mergers produce much more $^{56}$Ni, making them more akin to the sub-Chandrasekhar detonations also discussed in our work. As these uncertainties are more fully investigated, it may be worth revisiting our conclusions about the collision scenario.

Our conclusions do not rule out the collision mechanism for producing some fraction of SNe Ia. In fact, low luminosity 1991bg-like SNe Ia with $M_{\rm 56}\lesssim0.2M_\odot$ may be naturally explained by collisions in an old stellar environment, as shown in Figure \ref{fig:kushnir_mc} and discussed more extensively in \S \ref{sec:collisions}. This important connection should be explored by future investigations of this subclass of SNe Ia. 

\subsection{Missing Details and Future Work}

The investigation presented here uses a simple analysis to compare WD populations and explosion scenarios. Additional details should be included in future, more comprehensive calculations. For example, future similar work could use a more realistic star formation history \citep[for example, see][]{Ruiter et al. 2009} to explore the details of the resulting $^{56}$Ni distribution. In the Monte Carlo analysis, we used a bursty star formation rate set at various times in the past. This allowed us to demonstrate that a star formation rate more strongly peaked at earlier times would favor lower mass progenitors at later times since they take longer to evolve. {\em This naturally predicts lower luminosity SNe Ia in older populations because higher mass systems evolve more rapidly.} This may explain why late-type hosts have systematically brighter SNe Ia than early-type hosts, why the brightest events also occur in these kinds of galaxies \citep[e.g.,][]{Howell et al. 2007}, and why 1991bg-like SNe Ia happen almost exclusively in early-types galaxies. Detailed differences between early- and late-type may be an important tool for distinguishing between SNe Ia progenitor scenarios.

Another factor we have not completely accounted for is the timescale for detonation or collision in each scenario, and as a function of the WD masses. As long as $t_{\rm ign}$ is less than $t_{\rm form}$ (as we assumed in our work) this is a relatively small correction, but this need not be the case for all mass ratios. In particular, higher mass primaries have a wider range of possible companion masses. The ``eccentric Kozai mechanism'' (EKM), which promotes very strong eccentricity maxima and collisions in the inner binary of triple systems \citep{Ford et al. 2000,Naoz et al. 2011,Lithwick Naoz 2011,Katz et al. 2011,Naoz et al. 2013}, favors high mass ratio binaries and is suppressed over a wide range of tertiary inclinations when the masses of the inner binary are approximately equal \citep[see][]{Naoz et al. 2013,Shappee Thompson 2013}.  If EKM eccentricity maxima generically lead to collisions, then this would favor collisions in systems with higher $M_{\rm avg}$, which might help alleviate some of the inconsistencies seen in Figure \ref{fig:kushnir_mc}.  The EKM has also recently been shown to be enhanced over a broad range of parameter space in quadruple systems \citep{Pejcha et al. 2013}, potentially favoring WD-WD collisions in systems with initial mass distributions that might be different from normal binaries.

The machinery we have developed can be applied to new theoretical calculations of collisions and detonations, as well as to test other novel double degenerate scenarios. Some of the questions that would be particularly important to work out for inclusion in future calculations include the following.
\begin{itemize}
\item In collision scenarios, what is the $^{56}$Ni production as a function of the impact parameter and mass ratio?
\item In collision scenarios, how does the timescale for the collision ($t_{\rm ign}$ in our model) depend on the mass ratio?
\item If 1991bg-like SNe Ia are explained as collisions in old stellar environments, do their rates in late-type galaxies (which still have an old stellar component) match this hypothesis?
\item In sub-Chandrasekhar detonation scenarios, what is the expected $^{56}$Ni as a function of the $M_{\rm tot}$, and how does it depend on the mass ratio and time of ignition?
\item Extrapolating Figure \ref{fig:ni_yield} to high masses results in a large $^{56}$Ni yield for either scenario. As super-Chrandrasekhar SNe Ia are better characterized in comparison to regular SNe Ia, can these be naturally explained by either detonation or collision scenarios?
\end{itemize}
As these questions are better investigated, it should be worth revisiting and reevaluating many of the conclusions we have made here to gain a better understanding of what role double degenerates play in producing SNe Ia.

\acknowledgements
We thank Subo Dong for comments on a previous draft, which helped us improve our $^{56}$Ni yield estimates. We also thank Ashley Ruiter for discussions about population synthesis models and sharing her work for comparison. We thank Carles Badenes, Ryan Foley, Mohan Ganeshalingam, Peter Garnavich,  Saurabh Jha, Christian Ott, Ben Shappee, and Ken Shen for helpful feedback and assistance with interpreting observations. We also thank the Center for Cosmology and Astro-Particle Physics for funding ALP's visit to Ohio State University, where this work germinated.  ALP thanks John Beacom for generously arranging his visit. ALP is supported through NSF grants AST-1205732, PHY-1068881, PHY-1151197, and the Sherman Fairchild Foundation. TAT is supported in part by an NSF grant.


\begin{thebibliography}

\bibitem[Badenes 
\& Maoz(2012)]{Badenes Maoz 2012} Badenes, C., \& Maoz, D.\ 2012, \apjl, 749, L11 

\bibitem[Barbary et al.(2012)]{Barbary et al. 2012} Barbary, K., Aldering, 
G., Amanullah, R., et al.\ 2012, \apj, 745, 32 

\bibitem[Belczynski et al.(2008)]{Belcynski et al. 2008} Belczynski, K., 
Kalogera, V., Rasio, F.~A., et al.\ 2008, \apjs, 174, 223 

\bibitem[Bloom et al.(2012)]{Bloom et al. 2012} Bloom, J.~S., Kasen, D., 
Shen, K.~J., et al.\ 2012, \apjl, 744, L17 

\bibitem[Dan et al.(2012)]{Dan et al. 2012} Dan, M., Rosswog, S., 
Guillochon, J., \& Ramirez-Ruiz, E.\ 2012, \mnras, 422, 2417 

\bibitem[Fink et 
al.(2010)]{Fink et al. 2010} Fink, M., R{\"o}pke, F.~K., Hillebrandt, W., et al.\ 2010, \aap, 514, A53 

\bibitem[Folatelli et al.(2013)]{Folatelli et al. 2013} Folatelli, G., 
Morrell, N., Phillips, M.~M., et al.\ 2013, \apj, 773, 53 

\bibitem[Foley et al.(2013)]{Foley et al. 2013} Foley, R.~J., Challis, 
P.~J., Chornock, R., et al.\ 2013, \apj, 767, 57 

\bibitem[Ford et al.(2000)]{Ford et al. 2000} Ford, E.~B., Kozinsky, B., 
\& Rasio, F.~A.\ 2000, \apj, 535, 385 

\bibitem[Ganeshalingam et al.(2010)]{Ganeshalingam et al. 2010} Ganeshalingam, 
M., Li, W., Filippenko, A.~V., et al.\ 2010, \apjs, 190, 418 

\bibitem[Ganeshalingam et al.(2012)]{Ganeshalingam et al. 2012} Ganeshalingam, 
M., Li, W., Filippenko, A.~V., et al.\ 2012, \apj, 751, 142 

\bibitem[Gonz{\'a}lez-Gait{\'a}n et al.(2011)]{Gonzalez et al. 2011} 
Gonz{\'a}lez-Gait{\'a}n, S., Perrett, K., Sullivan, M., et al.\ 2011, \apj, 
727, 107 

\bibitem[Gonz{\'a}lez-Gait{\'a}n et al.(2012)]{Gonzalez et al. 2012} 
Gonz{\'a}lez-Gait{\'a}n, S., Conley, A., Bianco, F.~B., et al.\ 2012, \apj, 
745, 44 

\bibitem[Graur et al.(2011)]{Graur et al. 2011} Graur, O., Poznanski, D., 
Maoz, D., et al.\ 2011, \mnras, 417, 916 

\bibitem[Guillochon et al.(2010)]{gul10} Guillochon, J., Dan, 
M., Ramirez-Ruiz, E., \& Rosswog, S.\ 2010, \apjl, 709, L64

\bibitem[Hancock et al.(2011)]{Hancock et al. 2011} Hancock, P.~J., 
Gaensler, B.~M., \& Murphy, T.\ 2011, \apjl, 735, L35 

\bibitem[Hayden et al.(2010)]{Hayden et al. 2010} Hayden, B.~T., 
Garnavich, P.~M., Kasen, D., et al.\ 2010, \apj, 722, 1691 

\bibitem[Hicken et al.(2009)]{Hicken et al. 2009} Hicken, M., Challis, P., 
Jha, S., et al.\ 2009, \apj, 700, 331 

\bibitem[Horesh et al.(2012)]{Horesh et al. 2012} Horesh, A., Kulkarni, 
S.~R., Fox, D.~B., et al.\ 2012, \apj, 746, 21 

\bibitem[Howell et al.(2006)]{Howell et al. 2006} Howell, D.~A., Sullivan, 
M., Nugent, P.~E., et al.\ 2006, \nat, 443, 308 

\bibitem[Howell et al.(2007)]{Howell et al. 2007} Howell, D.~A., Sullivan, 
M., Conley, A., \& Carlberg, R.\ 2007, \apjl, 667, L37 

\bibitem[Hoyle 
\& Fowler(1960)]{hf60} Hoyle, F., \& Fowler, W.~A.\ 1960, \apj, 132, 565 

\bibitem[Iben 
\& Tutukov(1984)]{it84} Iben, I., Jr., \& Tutukov, A.~V.\ 1984, \apjs, 54, 335 

\bibitem[Kalirai et al.(2008)]{Kalirai et al. 2008} Kalirai, J.~S., Hansen, 
B.~M.~S., Kelson, D.~D., et al.\ 2008, \apj, 676, 594 

\bibitem[Kasen(2010)]{Kasen 2010} Kasen, D.\ 2010, \apj, 708, 1025 

\bibitem[Katz et al.(2011)]{Katz et al. 2011} Katz, B., Dong, S., 
\& Malhotra, R.\ 2011, Physical Review Letters, 107, 181101 

\bibitem[Katz 
\& Dong(2012)]{Katz Dong 2012} Katz, B., \& Dong, S.\ 2012, arXiv:1211.4584 

\bibitem[Kepler et al.(2007)]{Kepler et al. 2007} Kepler, S.~O., Kleinman, 
S.~J., Nitta, A., et al.\ 2007, \mnras, 375, 1315 

\bibitem[Khan et al.(2011)]{Khan et al. 2011} Khan, R., Stanek, K.~Z., 
Stoll, R., \& Prieto, J.~L.\ 2011, \apjl, 737, L24 

\bibitem[Krisciunas et al.(2000)]{Krisciunas et al. 2000} Krisciunas, K., 
Hastings, N.~C., Loomis, K., et al.\ 2000, \apj, 539, 658 

\bibitem[Krisciunas et al.(2004)]{Krisciunas et al. 2004} Krisciunas, K., 
Suntzeff, N.~B., Phillips, M.~M., et al.\ 2004, \aj, 128, 3034 

\bibitem[Kromer et al.(2010)]{Kromer et al. 2010} Kromer, M., Sim, S.~A., 
Fink, M., et al.\ 2010, \apj, 719, 1067 

\bibitem[Kushnir et al.~(2013)]{Kushnir et al. 2013} Kushnir, D., Katz, B., 
Dong, S., Livne, E., \& Fern{\'a}ndez, R.\ 2013, arXiv:1303.1180 

\bibitem[Leonard(2007)]{Leonard 2007} Leonard, D.~C.\ 2007, \apj, 
670, 1275 

\bibitem[Li et al.(2011a)]{Li et al. 2011a} Li, W., Bloom, J.~S., 
Podsiadlowski, P., et al.\ 2011a, \nat, 480, 348

\bibitem[Li et al.(2011b)]{Li et al. 2011b} Li, W., Leaman, J., 
Chornock, R., et al.\ 2011b, \mnras, 412, 1441 

\bibitem[Li et al.(2011c)]{Li et al. 2011c} Li, W., Chornock, R., 
Leaman, J., et al.\ 2011c, \mnras, 412, 1473 

\bibitem[Liebert et al.(2005)]{Liebert et al. 2005} Liebert, J., Bergeron, 
P., \& Holberg, J.~B.\ 2005, \apjs, 156, 47 

\bibitem[Lithwick 
\& Naoz(2011)]{Lithwick Naoz 2011} Lithwick, Y., \& Naoz, S.\ 2011, \apj, 742, 94 

\bibitem[Livne 
\& Arnett(1995)]{la95} Livne, E., \& Arnett, D.\ 1995, \apj, 452, 62

\bibitem[Mannucci et 
al.(2005)]{Mannucci et al. 2005} Mannucci, F., Della Valle, M., Panagia, N., et al.\ 2005, \aap, 433, 807 

\bibitem[Mannucci et al.(2006)]{Mannucci et al. 2006} Mannucci, F., Della 
Valle, M., \& Panagia, N.\ 2006, \mnras, 370, 773 

\bibitem[Maoz et al.(2010)]{Maoz et al. 2010} Maoz, D., Sharon, K., 
\& Gal-Yam, A.\ 2010, \apj, 722, 1879 

\bibitem[Mazzali et al.(2007)]{Mazzali et al. 2007} Mazzali, P.~A., 
R{\"o}pke, F.~K., Benetti, S., \& Hillebrandt, W.\ 2007, Science, 315, 825 

\bibitem[Modjaz et al.(2001)]{Modjaz et al. 2001} Modjaz, M., Li, W., 
Filippenko, A.~V., et al.\ 2001, \pasp, 113, 308 

\bibitem[Moore et al.(2013)]{Moore et al. 2013} Moore, K., Townsley, D., 
\& Bildsten, L.\ 2013, arXiv:1308.4193 

\bibitem[Naoz et al.(2011)]{Naoz et al. 2011} Naoz, S., Farr, W.~M., 
Lithwick, Y., Rasio, F.~A., \& Teyssandier, J.\ 2011, \nat, 473, 187

\bibitem[Naoz et al.(2013)]{Naoz et al. 2013} Naoz, S., Farr, W.~M., 
Lithwick, Y., Rasio, F.~A., \& Teyssandier, J.\ 2013, \mnras, 431, 2155

\bibitem[Nomoto 
\& Iben(1985)]{Nomoto Iben 1985} Nomoto, K., \& Iben, I., Jr.\ 1985, \apj, 297, 531 

\bibitem[Nomoto 
\& Kondo(1991)]{Nomoto Kondo 1991} Nomoto, K., \& Kondo, Y.\ 1991, \apjl, 367, L19 

\bibitem[Pakmor et al.(2010)]{Pakmor et al. 2010} Pakmor, R., Kromer, M., 
R{\"o}pke, F.~K., et al.\ 2010, \nat, 463, 61 

\bibitem[Pakmor et al.(2012)]{Pakmor et al. 2012} Pakmor, R., Kromer, M., 
Taubenberger, S., et al.\ 2012, \apjl, 747, L10

\bibitem[Pejcha et al.(2013)]{Pejcha et al. 2013} Pejcha, O., Antognini, 
J.~M., Shappee, B.~J., \& Thompson, T.~A.\ 2013, arXiv:1304.3152 

\bibitem[Perlmutter et al.(1999)]{per99} Perlmutter, S., 
Aldering, G., Goldhaber, G., et al.\ 1999, \apj, 517, 565 

\bibitem[Phillips(1993)]{Phillips 1993} Phillips, M.~M.\ 1993, \apjl, 
413, L105 

\bibitem[Piro 
\& Bildsten(2008)]{Piro Bildsten 2008} Piro, A.~L., \& Bildsten, L.\ 2008, \apj, 673, 1009 

\bibitem[Raghavan et al.(2010)]{Raghavan et al. 2010} Raghavan, D., 
McAlister, H.~A., Henry, T.~J., et al.\ 2010, \apjs, 190, 1 

\bibitem[Raskin et al.(2010)]{Raskin et al. 2010} Raskin, C., Scannapieco, 
E., Rockefeller, G., et al.\ 2010, \apj, 724, 111 

\bibitem[Riess et al.(1998)]{rie98} Riess, A.~G., Filippenko, 
A.~V., Challis, P., et al.\ 1998, \aj, 116, 1009 

\bibitem[Riess et al.(1999)]{rie99} Riess, A.~G., Filippenko, 
A.~V., Li, W., et al.\ 1999, \aj, 118, 2675 

\bibitem[Robitaille 
\& Whitney(2010)]{Robitaille Whitney 2010} Robitaille, T.~P., \& Whitney, B.~A.\ 2010, \apjl, 710, L11 

\bibitem[Rosswog et al.(2009)]{Rosswog et al. 2009} Rosswog, S., Kasen, D., 
Guillochon, J., \& Ramirez-Ruiz, E.\ 2009, \apjl, 705, L128 

\bibitem[Ruiter et al.(2009)]{Ruiter et al. 2009} Ruiter, A.~J., 
Belczynski, K., \& Fryer, C.\ 2009, \apj, 699, 2026 

\bibitem[Ruiter et al.(2013)]{Ruiter et al. 2013} Ruiter, A.~J., Sim, 
S.~A., Pakmor, R., et al.\ 2013, \mnras, 429, 1425 

\bibitem[Saio 
\& Nomoto(1998)]{Saio Nomoto 1998} Saio, H., \& Nomoto, K.\ 1998, \apj, 500, 388 

\bibitem[Sand et al.(2012)]{Sand et al. 2012} Sand, D.~J., Graham, 
M.~L., Bildfell, C., et al.\ 2012, \apj, 746, 163 

\bibitem[Scannapieco 
\& Bildsten(2005)]{Scannapieco Bildsten 2005} Scannapieco, E., \& Bildsten, L.\ 2005, \apjl, 629, L85 

\bibitem[Schaefer 
\& Pagnotta(2012)]{Schaefer Pagnotta 2012} Schaefer, B.~E., \& Pagnotta, A.\ 2012, \nat, 481, 164 

\bibitem[Shappee et al.(2013a)]{Shappee et al. 2013a} Shappee, B.~J., Stanek, 
K.~Z., Pogge, R.~W., \& Garnavich, P.~M.\ 2013a, \apjl, 762, L5 

\bibitem[Shappee et al.(2013b)]{Shappee et al. 2013b} Shappee, B.~J., 
Kochanek, C.~S., \& Stanek, K.~Z.\ 2013b, \apj, 765, 150 

\bibitem[Shappee 
\& Thompson(2013)]{Shappee Thompson 2013} Shappee, B.~J., \& Thompson, T.~A.\ 2013, \apj, 766, 64 

\bibitem[Shen et al.(2012)]{Shen et al. 2012} Shen, K.~J., Bildsten, L., 
Kasen, D., \& Quataert, E.\ 2012, \apj, 748, 35 

\bibitem[Sim et al.(2010)]{Sim et al. 2010} Sim, S.~A., R{\"o}pke, 
F.~K., Hillebrandt, W., et al.\ 2010, \apjl, 714, L52 

\bibitem[Stritzinger et 
al.(2006)]{Stritzinger et al. 2006} Stritzinger, M., Mazzali, P.~A., Sollerman, J., \& Benetti, S.\ 2006, \aap, 460, 793 

\bibitem[Sullivan et al.(2006)]{Sullivan et al. 2006} Sullivan, M., Le 
Borgne, D., Pritchet, C.~J., et al.\ 2006, \apj, 648, 868 

\bibitem[Sullivan et al.(2011)]{Sullivan et al. 2011} Sullivan, M., 
Kasliwal, M.~M., Nugent, P.~E., et al.\ 2011, \apj, 732, 118 

\bibitem[Thompson(2011)]{Thompson 2011} Thompson, T.~A.\ 2011, \apj, 
741, 82 

\bibitem[Townsley et al.(2012)]{Townsley et al. 2012} Townsley, D.~M., 
Moore, K., \& Bildsten, L.\ 2012, \apj, 755, 4 

\bibitem[van Kerkwijk et al.(2010)]{van Kerkwijk et al. 2010} van Kerkwijk, 
M.~H., Chang, P., \& Justham, S.\ 2010, \apjl, 722, L157 

\bibitem[Vennes(1999)]{Vennes 1999} Vennes, S.\ 1999, \apj, 525, 
995 

\bibitem[Wang et al.(2009)]{Wang et al. 2009} Wang, X., Filippenko, 
A.~V., Ganeshalingam, M., et al.\ 2009, \apjl, 699, L139 

\bibitem[Webbink(1984)]{web84} Webbink, R.~F.\ 1984, \apj, 
277, 355 

\bibitem[Wegg 
\& Phinney(2012)]{Wegg Phinney 2012} Wegg, C., \& Phinney, E.~S.\ 2012, \mnras, 426, 427

\bibitem[Whelan 
\& Iben(1973)]{wi73} Whelan, J., \& Iben, I., Jr.\ 1973, \apj, 186, 1007 

\bibitem[Wickramasinghe 
\& Ferrario(2000)]{Wickramasinghe Ferrario 2000} Wickramasinghe, D.~T., \& Ferrario, L.\ 2000, \pasp, 112, 873 

\bibitem[Woosley 
\& Weaver(1994a)]{ww94} Woosley, S.~E., \& Weaver, T.~A.\ 1994, \apj, 423, 371 

\end{thebibliography}
\end{document}